\documentclass[conference]{IEEEtran}

\usepackage{amsfonts}
\usepackage{amsmath}
\usepackage{amssymb}
\usepackage{lscape}
\usepackage{epsf}
\usepackage{graphicx}
\usepackage{psfrag}

\newtheorem{theorem}{\indent Theorem}[section]
\newtheorem{lemma}[theorem]{\indent Lemma}

\newtheorem{EXAMPLE}{\indent Example}[section]
\newtheorem{definition}{\indent Definition}[section]

\newcommand{\cA}{{\mathcal{A}}}

\newcommand{\cB}{{\mathcal{B}}}

\newcommand{\cM}{{\mathcal{M}}}

\newcommand{\coeff}{{\mbox{Coeff }}}

\newcommand{\bldbeta}{{\mbox{\boldmath $\beta$}}}
\newcommand{\bldsmallbeta}{{\mbox{\scriptsize \boldmath $\beta$}}}

\newcommand{\bldeta}{{\mbox{\boldmath $\eta$}}}
\newcommand{\bldsmalleta}{{\mbox{\scriptsize \boldmath $\eta$}}}
\newcommand{\bldepsilon}{{\mbox{\boldmath $\epsilon$}}}

\newcommand{\bldgamma}{{\mbox{\boldmath $\gamma$}}}
\newcommand{\bldsmallgamma}{{\mbox{\scriptsize \boldmath $\gamma$}}}

\newcommand{\bldnu}{{\mbox{\boldmath $\nu$}}}
\newcommand{\bldsmallnu}{{\mbox{\scriptsize \boldmath $\nu$}}}

    \def\squarebox#1{\hbox to #1{\hfill\vbox to #1{\vfill}}}

\newlength{\Algwidth}

\title{On the Growth Rate of the Weight Distribution of Irregular Doubly-Generalized LDPC Codes}

\author{
\authorblockN{Mark F. Flanagan, Enrico Paolini and Marco Chiani}
\authorblockA{DEIS, University of Bologna \\
via Venezia 52, 47023 Cesena (FC), Italy  \\
Email: mark.flanagan@ieee.org, e.paolini@unibo.it, marco.chiani@unibo.it}
\authorblockN{Marc Fossorier}
\authorblockA{ETIS ENSEA / UCP / CNRS UMR-8051 \\
6, avenue du Ponceau, 95014 Cergy Pontoise, France \\
Email: mfossorier@ieee.org}
}

\begin{document}
\maketitle

\begin{abstract}
In this paper, an expression for the asymptotic growth rate of the
number of small linear-weight codewords of irregular doubly-generalized
LDPC (D-GLDPC) codes is derived. The expression is compact and generalizes 
existing results for LDPC and generalized LDPC (GLDPC) codes. Assuming that there exist check and variable nodes with minimum distance $2$, it is shown that the growth rate depends only on these nodes. An important
connection between this new result and the stability condition of D-GLDPC codes
over the BEC is highlighted. Such a connection, previously
observed for LDPC and GLDPC codes, is now extended to
the case of D-GLDPC codes.
\end{abstract}
\begin{keywords}
Doubly-generalized LDPC codes,
irregular code ensembles,
weight distribution. 
\end{keywords}

\section{Introduction}

Recently, low-density parity-check (LDPC) codes have been intensively studied due to
their near-Shannon-limit performance under iterative
belief-propagation decoding. Binary regular LDPC codes were first
proposed by Gallager in 1963 \cite{Gallager}. In the last
decade the capability of irregular LDPC codes to outperform
regular ones in the waterfall region of the performance curve and
to asymptotically approach (or even achieve) the communication
channel capacity has been recognized and deeply investigated (see
for instance
\cite{luby01:improved,luby01:efficient,richardson01:design,richardson01:dB,pfister05:capacity-achieving,pfister07:ara}).

It is usual to represent an LDPC code as a bipartite graph,
i.e., as a graph where the nodes are grouped into two disjoint
sets, namely, the variable nodes (VNs) and the check nodes (CNs), such that each
edge may only connect a VN to a CN. The bipartite graph
is also known as a Tanner graph \cite{Tanner_GLDPC}. In the Tanner
graph of an LDPC code, a generic degree-$q$ VN can
be interpreted as a length-$q$ repetition code, as it repeats $q$
times its single information bit towards the CNs. Similarly, a degree-$s$ 
CN of an LDPC code can be interpreted
as a length-$s$ single parity-check (SPC) code, as it checks the parity of the
$s$ VNs connected to it.

The growth rate of the weight distribution of Gallager's regular
LDPC codes was investigated in \cite{Gallager}. The analysis
demonstrated that, provided that the smallest
VN degree is at least 3, for large enough codeword length $N$, the expected minimum distance of a randomly chosen code in the ensemble is a linear
function of $N$.

More recently, the study of the weight distribution of binary LDPC codes
has been extended to irregular ensembles. Important works in this
area are
\cite{litsyn02:ensembles,Burshtein_Miller,Di_Richardson_Urbanke}.
In \cite{Di_Richardson_Urbanke} a complete solution for the growth rate of the weight
distribution of binary irregular LDPC codes was
developed. One of the main results of \cite{Di_Richardson_Urbanke} is a connection
between the expected behavior of the weight distribution of a code
randomly chosen from the ensemble and the parameter
$\lambda'(0)\rho'(1)$, $\lambda(x)$ and $\rho(x)$ being the
edge-perspective VN and CN degree distributions,
respectively. More specifically, it was shown that for a code randomly
chosen from the ensemble, one can expect an exponentially small
number of small linear-weight codewords if $0 \leq
\lambda'(0)\rho'(1)<1$, and an exponentially large number of small
linear-weight codewords if $\lambda'(0)\rho'(1)>1$.

This result establishes a connection between the statistical
properties of the weight distribution of binary irregular LDPC codes and
the stability condition of binary irregular LDPC codes over the binary erasure channel (BEC)
\cite{luby01:efficient,richardson01:design}. If $q^*$ denotes
the LDPC asymptotic iterative decoding threshold over the
BEC, the stability condition states that we always have
\begin{align}\label{eq:stability_LDPC}
q^* \leq \left[ \lambda'(0)\rho'(1) \right]^{-1}.
\end{align}

Prior to the rediscovery of LDPC codes, binary generalized LDPC (GLDPC) codes
were introduced by Tanner in 1981 \cite{Tanner_GLDPC}. A
GLDPC code generalizes the concept of an LDPC code in that
a degree-$s$ CN may in principle be any $(s,h)$ linear block
code, $s$ being the code length and $h$ the code dimension. Such a
CN accounts for $s-h$ linearly independent parity-check equations. A CN
associated with a linear block code which is not a SPC code
is said to be a \emph{generalized CN}. In \cite{Tanner_GLDPC}
\emph{regular} GLDPC codes (also known as Tanner codes) were
investigated, these being GLDPC codes where the VNs are
all repetition codes of the same length and the
CNs are all linear block codes of the same type.

\begin{figure*}[t]
\begin{center}
\psfrag{n2edges}{\small{$s$ edges}} \psfrag{k1bits}{ \small{$k$
bits}} \psfrag{n2k2genCN}{$(s,h)$ generalized CN}
\psfrag{n2k2eqns}{\small{$s-h$ equations}} \psfrag{SPCCN}{SPC CN}
\psfrag{repVN}{rep. VN} \psfrag{n1k1genVN}{$(q,k)$ generalized VN}
\psfrag{brace}{$\underbrace{\phantom{---------------------------}}$}
\psfrag{encodedbits}{$N$ code bits} \psfrag{n1edges}{\small{$q$
edges}}
\includegraphics[width=6 cm, angle=270]{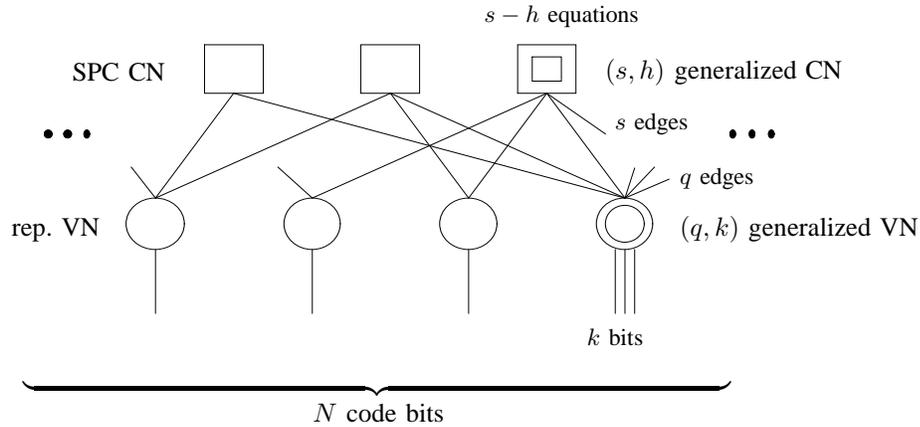}
\end{center}
\caption{Structure of a D-GLDPC code.} \label{fig:DGLDPC}
\end{figure*}

The growth rate of the weight distribution of binary GLDPC codes was
investigated in
\cite{boutros99:generalized,lentmaier99:generalized,Tillich04:weight,paolini08:weight}.
In \cite{boutros99:generalized} the growth rate is calculated for
Tanner codes with BCH check component codes and length-2
repetition VNs, leading to an asymptotic lower bound on the
minimum distance. The same lower bound is developed in
\cite{lentmaier99:generalized} assuming Hamming CNs and length-2 repetition VNs. Both works extend
the approach developed by Gallager in \cite[Chapter 2]{Gallager}
to show that, for sufficiently large $N$, the minimum distance is a
linear function of $N$. The growth rate of the number of small
weight codewords for GLDPC codes with a uniform CN set (all CN of the same type) and an irregular VN set
(repetition VNs with different lengths) is investigated in \cite{Tillich04:weight}. It
is shown that for sufficiently large $N$, a minimum distance 
increasing linearly with $N$ is expected when either the uniform CN set
is composed of linear block codes with minimum distance at least $3$, or the
minimum length of the repetition VNs is 3. On the other hand,
if the minimum distance of the CNs and the minimum length of the
repetition VNs are both equal to 2, then for a randomly
selected GLDPC code in the ensemble we expect a minimum
distance growing as a linear or sublinear function of $N$ (for
large $N$) depending on the sign of the first order coefficient in
the growth rate Taylor series expansion. The results developed in
\cite{Tillich04:weight} were further extended in
\cite{paolini08:weight} to GLDPC ensembles with an irregular
CN set (CNs of different types). It was there
proved that, provided that there exist CNs with minimum distance $2$, a
parameter $\lambda'(0)C$, generalizing the parameter
$\lambda'(0)\rho'(1)$ of LDPC code ensembles, plays in the context
of the weight distribution of GLDPC codes the same role played by
$\lambda'(0)\rho'(1)$ in the context of the weight
distribution of LDPC codes. The parameter $C$ is defined in
Section~\ref{section:further_def_&_notation}.

Interestingly, this latter results extends to binary GLDPC codes the
same connection between the statistical properties of the weight distribution of irregular
codes and the stability condition over the
BEC. In fact, it was shown in \cite{paolini08:stability} that
the stability condition of binary irregular GLDPC codes over the
BEC is given by
\begin{align}\label{eq:stability_GLDPC}
q^* \leq \left[ \lambda'(0)C \right]^{-1}.
\end{align}

Generalized LDPC codes represent a promising solution for
low-rate channel coding schemes, due to an overall rate loss
introduced by the generalized CNs
\cite{miladinovic08:generalized}. Doubly-generalized LDPC (D-GLDPC) codes
generalize the concept of GLDPC codes while facilitating much greater design 
flexibility in terms of code rate \cite{Wang_Fossorier_DG_LDPC} (an analogous idea may be
found in the previous work \cite{Dolinar}). In a D-GLDPC code,
the VNs as well as the CNs may be of any generic linear block code
types. A degree-$q$ VN may in principle be any $(q,k)$ linear
block code, $q$ being the code length and $k$ the code dimension.
Such a VN is associated with $k$ D-GLDPC code bits. It interprets these bits as its local information bits and interfaces
to the CN set through its $q$ local code bits. A VN which corresponds to a 
linear block code which is not a repetition code is said to be a
\emph{generalized VN}. A D-GLDPC code is
said to be \emph{regular} if all of its VNs are of the same type
and all of its CNs are of the same type and is said to be \emph{irregular}
otherwise. The structure of a D-GLDPC code is depicted in
Fig.~\ref{fig:DGLDPC}.

In this paper the growth rate of the weight distribution of binary irregular D-GLDPC codes
is analyzed for small weight codewords. It is
shown that, provided there exist both VNs
and CNs with minimum distance $2$, a parameter $1/P^{-1}(1/C)$
discriminates between an asymptotically small and an
asymptotically large expected number of small linear-weight
codewords in a D-GLDPC code randomly drawn from a given
irregular ensemble (the function $P(x)$ is defined in
Section~\ref{section:further_def_&_notation}). The parameter
$1/P^{-1}(1/C)$ generalizes the above mentioned parameters
$\lambda'(0)\rho'(1)$ and $\lambda'(0)C$ to the case where both
generalized VNs and generalized CNs are present. The
obtained result also represents the extension to the D-GLDPC
case of the previously recalled connection with the stability
condition over the BEC. In fact, it was proved in
\cite{paolini08:stability} that the stability condition of
D-GLDPC codes over the BEC is given by
\begin{align}\label{eq:stability_D-GLDPC}
q^* \leq P^{-1}(1/C)\, .
\end{align}
The paper is organized as follows. Section \ref{section:irregular_D_GLDPC} defines the D-GLDPC ensemble of interest, and introduces some definitions and notation pertaining to this ensemble. Section~\ref{section:further_def_&_notation} defines further terms regarding the VNs and CNs which compose the D-GLDPC codes in the ensemble. Finally, Section~\ref{section:growth_rate} states and proves the main result of the paper regarding the growth rate of the weight distribution.

\section{Irregular Doubly-Generalized LDPC Code Ensemble}
\label{section:irregular_D_GLDPC}

We define a D-GLDPC code ensemble $\cM_n$ as follows, where $n$ denotes the number of VNs. There are $n_c$ different CN types $t \in I_c = \{ 1,2,\cdots, n_c\}$, and $n_v$ different VN types $t \in I_v = \{ 1,2,\cdots, n_v\}$. For each CN type $t \in I_c$, we denote by $h_t$, $s_t$ and $r_t$ the CN dimension, length and minimum distance, respectively. For each VN type $t \in I_v$, we denote by $k_t$, $q_t$ and $p_t$ the VN dimension, length and minimum distance, respectively. For $t \in I_c$, $\rho_t$ denotes the fraction of edges connected to CNs of type $t$. Similarly, for $t \in I_v$, $\lambda_t$ denotes the fraction of edges connected to VNs of type $t$. Note that all of these variables are independent of $n$.

The polynomials $\rho(x)$ and $\lambda(x)$ are defined by
\[
\rho(x) = \sum_{t\in I_c} \rho_t x^{s_t - 1}
\]   
and
\[
\lambda(x) = \sum_{t \in I_v} \lambda_t x^{q_t - 1} \; .
\]   
If $E$ denotes the number of edges in the Tanner graph, the number of CNs of type $t\in I_c$ is then given by $E \rho_t / s_t$, and the number of VNs of type $t\in I_v$ is then given by $E \lambda_t / q_t$. Denoting as usual $\int_0^1 \rho(x) \, {\rm d} x$ and $\int_0^1 \lambda(x) \, {\rm d} x$ by $\int \rho$ and $\int \lambda$ respectively, we see that the number of edges in the Tanner graph is given by
\[
E = \frac{n}{\int \lambda}
\]
and the number of CNs is given by $m = E \int \rho$. Therefore, the fraction of CNs of type $t \in I_c$ is given by
\begin{equation}
\gamma_t = \frac{\rho_t}{s_t \int \rho}
\label{eq:gamma_t_definition}
\end{equation}
and the fraction of VNs of type $t \in I_v$ is given by
\begin{equation}
\delta_t = \frac{\lambda_t}{q_t \int \lambda}
\label{eq:delta_t_definition}
\end{equation}
Also the length of any D-GLDPC codeword in the ensemble is given by 
\begin{equation}
N = \sum_{t \in I_v} \left( \frac{E \lambda_t}{q_t} \right) k_t = \frac{n}{\int \lambda} \sum_{t \in I_v} \frac{\lambda_t k_t}{q_t} \; .
\label{eq:DG_LDPC_codeword_length}
\end{equation}
Note that this is a linear function of $n$. Similarly, the total number of parity-check equations for any D-GLDPC code in the ensemble is given by
\[
M = \frac{m}{\int \rho} \sum_{t \in I_c} \frac{\rho_t h_t}{s_t} \; .
\]
A member of the ensemble then corresponds to a permutation on the $E$ edges connecting CNs to VNs.

The growth rate of the weight distribution of the irregular D-GLDPC ensemble sequence $\{ \cM_n \}$ is defined by 
\begin{equation}
G(\alpha) = \lim_{n\rightarrow \infty} \frac{1}{n} \log \mathbb{E}_{\cM_n} \left[ N_{\alpha n} \right]
\label{eq:growth_rate_result}
\end{equation}
where $\mathbb{E}_{\cM_n}$ denotes the expectation operator over the ensemble $\cM_n$, and $N_{w}$ denotes the number of codewords of weight $w$ of a randomly chosen D-GLDPC code in the ensemble $\cM_n$. The limit in (\ref{eq:growth_rate_result}) assumes the inclusion of only those positive integers $n$ for which $\alpha n \in \mathbb{Z}$ and $\mathbb{E}_{\cM_n} [ N_{\alpha n} ]$ is positive (i.e., where the expression whose limit we seek is well defined). Note that the argument of the growth rate function $G(\alpha)$ is equal to the ratio of D-GLDPC codeword length to the number of VNs; by (\ref{eq:DG_LDPC_codeword_length}), this captures the behaviour of codewords linear in the block length, as in \cite{Di_Richardson_Urbanke} for the LDPC case. 
\medskip
\begin{definition}
An \emph{assignment} is a subset of the edges of the Tanner graph. An assignment is said to have \emph{weight} $k$ if it has $k$ elements. An assignment is said to be \emph{check-valid} if the following condition holds: supposing that each edge of the assignment carries a $1$ and each of the other edges carries a $0$, each CN recognizes a valid codeword. 
\end{definition}
\medskip
\begin{definition}
A \emph{split assignment} is an assignment, together with a subset of the D-GLDPC code bits (called a \emph{codeword assignment}). A split assignment is said to have \emph{split weight} $(u, v)$ if its assignment has weight $v$ and its codeword assignment has $u$ elements. A split assignment is said to be \emph{check-valid} if its assignment is check-valid. A split assignment is said to be \emph{variable-valid} if the following condition holds: supposing that each edge of its assignment carries a $1$ and each of the other edges carries a $0$, and supposing that each D-GLDPC code bit in the codeword assigment is set to $1$ and each of the other code bits is set to $0$, each VN recognizes an input word and the corresponding valid codeword.     
\end{definition}
\medskip
Note that for any D-GLDPC code, there is a bijective correspondence between the set of D-GLDPC codewords and the set of split assignments which are both variable-valid and check-valid. 
\section{Further Definitions and Notation}
\label{section:further_def_&_notation}
The weight enumerating polynomial for CN type $t \in I_c$ is given by 
\begin{eqnarray*}
A^{(t)}(x) & = & \sum_{u=0}^{s_t} A_u^{(t)} x^u \\
& = & 1 + \sum_{u=r_t}^{s_t} A_u^{(t)} x^u \; .
\end{eqnarray*}
Here $A_u^{(t)} \ge 0$ denotes the number of weight-$u$ codewords for CNs of type $t$. Note that $A_{r_t}^{(t)} > 0$ for all $t \in I_c$. The bivariate weight enumerating polynomial for VN type $t \in I_v$ is given by 
\begin{eqnarray*}
B^{(t)}(x,y) & = & \sum_{u=0}^{k_t} \sum_{v=0}^{q_t} B_{u,v}^{(t)} x^u y^v \\
& = & 1 + \sum_{u=1}^{k_t} \sum_{v=p_t}^{q_t} B_{u,v}^{(t)} x^u y^v \; .
\end{eqnarray*}
Here $B_{u,v}^{(t)} \ge 0$ denotes the number of weight-$v$ codewords generated by input words of weight $u$, for VNs of type $t$. Also, for each $t \in I_v$, corresponding to the polynomial $B^{(t)}(x,y)$ we denote the sets
\begin{equation}
S_t = \{ (i,j) \in \mathbb{Z}^2 \; : \; B^{(t)}_{i,j} > 0 \}
\label{eq:St}
\end{equation}
and
\begin{equation}
S_t^{-} = S_t \backslash \{ (0,0) \} \; .
\label{eq:St-}
\end{equation}

We denote the smallest minimum distance over all CN types by
\[
r = \min \{ r_t \; : \; t \in I_c \}
\]
and the set of CN types with this minimum distance by
\[
X_c = \{ t \in I_c \; : \; r_t = r \} \; .
\]
We also define
\[
C_t = \frac{r_t A^{(t)}_{r_t}}{s_t}
\] 
for each $t \in I_c$, and
\begin{equation}
C = \sum_{t \in X_c} \rho_t C_t \; .
\label{eq:C_definition}
\end{equation}
We note that $C_t>0$ for all $t \in I_c$, so $C > 0$.
 
Similarly, we denote the smallest minimum distance over all VN types by
\[
p = \min \{ p_t \; : \; t \in I_v \}
\]
and the set of VN types with this minimum distance by 
\[
X_v = \{ t \in I_v \; : \; p_t = p \} \; .
\]
In the specific case where $p = 2$, we also introduce the following definitions. For each $t \in X_v$, define the set $L_t = \{ i \in \mathbb{Z} \; : \; B^{(t)}_{i,2} > 0 \}$ -- note that these sets are nonempty. Also define the polynomial $P(x)$ by
\begin{equation}
P(x) = \sum_{t \in X_v} \lambda_t \sum_{i \in L_t} \frac{2 B^{(t)}_{i,2}}{q_t} x^i
\label{eq:Px_definition}
\end{equation}
and denote its inverse by $P^{-1}(x)$. Since all the coefficients of $P(x)$ are positive, $P(x)$ is monotonically increasing and therefore its inverse is well-defined and unique. Note that in the case $r=p=2$, both $C$ and the polynomial $P(x)$ depend only on the CNs and VNs with minimum distance equal to $2$.

Finally, note that throughout this paper, the notation $e = \exp(1)$ denotes Napier's number. 

\section{Growth Rate for Doubly-Generalized LDPC Code Ensemble}
\label{section:growth_rate}
The following theorem constitutes the main result of the paper.
\medskip
\begin{theorem}
Consider an irregular D-GLDPC code ensemble sequence $\cM_n$ satisfying $r=p=2$. For sufficiently small $\alpha$, the growth rate of the weight distribution is given by
\begin{equation}
G(\alpha) = \alpha \log \left[ \frac{1}{P^{-1}(1/C)} \right] + O(\alpha^2) \; .
\label{eq:growth_rate_case_1}
\end{equation}
\label{thm:growth_rate} 
\end{theorem}
The theorem is proved next. For ease of presentation, the proof is broken into four parts.

\subsection{Number of check-valid assignments of weight $\epsilon m$ over $\gamma m$ CNs of type $t \in I_c$}
Consider $\gamma m$ CNs of the same type $t \in I_c$. Using generating functions
\footnote{Here we make use of the following general result \cite{Wilf}. Let $a_i$ be the number of ways of obtaining an outcome $i\in\mathbb{Z}$ in experiment $\cA$, and let $b_j$ be the number of ways of obtaining an outcome $j\in\mathbb{Z}$ in experiment $\cB$. Also let $c_k$ be the number of ways of obtaining an outcome $(i,j)$ in the combined experiment $(\cA, \cB)$ with sum $i+j=k$. Then the generating functions $A(x)=\sum_i a_i x^i$, $B(x)=\sum_j b_j x^j$ and $C(x)=\sum_k c_k x^k$ are related by $C(x) = A(x) B(x)$.}
, the number of check-valid assignments (over these CNs) of weight $\epsilon m$ is given by 
\[
N_{c,t}^{(\gamma m)}(\epsilon m) = \coeff \left[ \left( A^{(t)}(x) \right) ^{\gamma m}, x^{\epsilon m} \right]
\]
where $\coeff [ p(x), x^c ]$ denotes the coefficient of $x^c$ in the polynomial $p(x)$. We now use the following result, which appears as Lemma 19 in \cite{Di_Richardson_Urbanke}:
\medskip
\begin{lemma}
Let $A(x) = 1 + \sum_{u=c}^{d} A_u x^u$, where $1 \le c \le d$, be a polynomial satisfying $A_c > 0$ and $A_u \ge 0$ for all $c < u \le d$. Then, for sufficiently small $\xi$, 
\begin{multline}
\lim_{\ell\rightarrow \infty} \frac{1}{\ell} \log \coeff \left[ \left( A(x) \right) ^{\ell}, x^{\xi \ell} \right] \\
= \frac{\xi}{c} \log \left( \frac{e c A_c}{\xi} \right) + O(\xi^2)
\label{lemma_42_formula}
\end{multline}
\label{lemma:optimization_dominant_term_1D}
\end{lemma}
\medskip
The limit in (\ref{lemma_42_formula}) assumes the inclusion of only those positive integers $\ell$ for which $\xi \ell \in \mathbb{Z}$ and $\coeff [ ( A(x) ) ^{\ell}, x^{\xi \ell} ]$ is positive (i.e., where the expression whose limit we seek is well defined). A proof of this lemma may be found in the Appendix; our proof is based on arguments from \cite{Burshtein_Miller} and Lagrange multipliers, and constitutes a different approach to that taken in \cite{Di_Richardson_Urbanke}.

Applying this lemma by substituting $A(x) = A^{(t)}(x)$, $\ell=\gamma m$ and $\xi = \epsilon/\gamma$, we obtain that with $\gamma$ fixed, as $m \rightarrow \infty$ we have, for sufficiently small $\epsilon$, 
\begin{multline}
N_{c,t}^{(\gamma m)}(\epsilon m) = \coeff \left[ \left( A^{(t)}(x) \right) ^{\gamma m}, x^{\epsilon m} \right] \rightarrow \\
\exp \left\{ m \left[ \frac{\epsilon}{r_t} \log \left( \frac{e r_t A^{(t)}_{r_t} \gamma}{\epsilon} \right) + O(\epsilon^2) \right] \right\}
\label{eq:check_node_type_t}
\end{multline}
\subsection{Number of check-valid assignments of weight $\delta m$}
\label{sub:no_check_valid_assign}
Next we derive an expression, valid asymptotically, for the number of check-valid assignments of weight $\delta m$. For each $t \in I_c$, let $\epsilon_t m$ denote the portion of the total weight $\delta m$ apportioned to CNs of type $t$. Then $\epsilon_t \ge 0$ for each $t \in I_c$, and $\sum_{t \in I_c} \epsilon_t = \delta$. Also denote $\bldepsilon = (\epsilon_1 \; \epsilon_2 \; \cdots \; \epsilon_{n_c})$. The number of check-valid assignments of weight $\delta m$ satisfying the constraint $\bldepsilon$ is obtained by multiplying the numbers of check-valid assignments of weight $\epsilon_t m$ over $\gamma_t m$ CNs of type $t$, for each $t \in I_c$,
\[
N_c^{(\bldepsilon)}(\delta m) = \prod_{t \in I_c} N_{c,t}^{(\gamma_t m)}(\epsilon_t m) 
\]
where the fraction $\gamma_t$ of CNs of type $t \in I_c$ is given by (\ref{eq:gamma_t_definition}).

As $n\rightarrow \infty$, we have $m\rightarrow \infty$ and so we obtain using (\ref{eq:check_node_type_t}) that for sufficiently small $\delta$, 
\begin{multline}
N_c^{(\bldepsilon)}(\delta m) \rightarrow \\ 
\prod_{t \in I_c} \exp \left\{ m \left[ \frac{\epsilon_t}{r_t} \log \left( \frac{e r_t A^{(t)}_{r_t} \gamma_t}{\epsilon_t} \right) + O(\epsilon_t^2) \right] \right\}  \\
= \exp \left\{ m \left[ \sum_{t \in I_c} \left( \frac{\epsilon_t}{r_t} \log \left( \frac{e \rho_t C_t}{\epsilon_t \int \rho} \right) \right) + O(\delta^2) \right] \right\} 
\label{eq:number_of_assign_distribution}
\end{multline}

The number of check-valid assignments of weight $\delta m$, which we denote $N_c(\delta m)$, is equal to the sum of $N_c^{(\bldepsilon)}(\delta m)$ over all admissible vectors $\bldepsilon$. However, the asymptotic expression as $n\rightarrow \infty$ will be dominated by the distribution $\bldepsilon$ which maximizes the argument of the exponential
\footnote{Observe that as $m\rightarrow \infty$, $\sum_t \exp ( m Z_t ) \rightarrow \exp ( m \max_t \{Z_t\} )$}
. Therefore, our next step is to maximize the function  
\[
f(\bldepsilon) = \sum_{t \in I_c} \frac{\epsilon_t}{r_t} \log \left( \frac{e \rho_t C_t}{\epsilon_t \int \rho} \right)
\]
subject to the constraints 
\begin{equation}
g(\bldepsilon) = \sum_{t \in I_c} \epsilon_t = \delta 
\label{eq:sum_epsilon_equals_1}
\end{equation}
and 
\begin{equation}
\epsilon_t \ge 0 \quad \forall t \in I_c \; .
\label{eq:nonnegative_constraint_epsilon}
\end{equation}
We solve this optimization problem using Lagrange multipliers, ignoring for the moment the final constraint. Since
\[
\frac{\partial f}{\partial\epsilon_{t}} = \frac{1}{r_t} \log \left( \frac{\rho_t C_t}{\epsilon_t \int \rho} \right) \quad ; \quad \frac{\partial g}{\partial\epsilon_t} = 1
\]
for all $t \in I_c$, we have to solve the $n_c$ equations (where $\mu$ is the Lagrange multiplier)
\begin{equation}
\frac{1}{r_t} \log \left( \frac{\rho_t C_t}{\epsilon_t \int \rho} \right) + \mu = 0 \quad \forall t \in I_c
\label{eq:Lagrange_epsilons}
\end{equation}
together with (\ref{eq:sum_epsilon_equals_1}), for the $(n_c+1)$ unknowns $\{ \mu, \bldepsilon \}$. First, (\ref{eq:Lagrange_epsilons}) yields 
\[
\epsilon_t = \frac{\rho_t C_t}{\int \rho} z^{r_t} \quad \forall t \in I_c
\]
where $z = e^{\mu}$. Now using (\ref{eq:sum_epsilon_equals_1}), we obtain
\[
\frac{1}{\int \rho} \sum_{t \in I_c} C_t \rho_t z^{r_t} = \delta \; .
\]
The left-hand side of this equation involves a sum of positive terms. For $\delta$ sufficiently small, we may approximate 
\begin{eqnarray}
\frac{1}{\int \rho} \sum_{t \in I_c} \rho_t C_t z^{r_t} & \approx & \frac{1}{\int \rho} \sum_{t \in X_c} \rho_t C_t z^{r_t} \\
& = & \frac{C}{\int \rho} z^r \; .
\end{eqnarray}
Applying this approximation, we obtain $\epsilon_t = 0$ if $t \notin X_c$, and $\epsilon_t = K \rho_t C_t$ if $t \in X_c$, where $K$ is independent of $t$. Then (\ref{eq:sum_epsilon_equals_1}) yields $K = \delta / C$, and we obtain the solution  
\[
\epsilon_t = \left\{ \begin{array}{cl}
\rho_t C_t \delta / C & \textrm{if } t \in X_c \\
0 & \textrm{otherwise.}\end{array}\right.
\]
This solution satisfies (\ref{eq:nonnegative_constraint_epsilon}). When substituted into (\ref{eq:number_of_assign_distribution}), it yields the following result: as $n\rightarrow \infty$  
\begin{equation}
N_c(\delta m) \rightarrow \exp \left\{ m \left[ \frac{\delta}{r} \log \left( \frac{e C}{\delta \int \rho} \right) + O(\delta^2) \right] \right\}
\label{eq:number_of_check_valid_assign} 
\end{equation}

\subsection{Number of variable-valid split assignments of split weight $(\tau n, \sigma n)$ over $\gamma n$ VNs of type $t \in I_v$}
Consider $\gamma n$ VNs of the same type $t \in I_v$. We now evaluate the number of variable-valid split assignments (over these VNs) of split weight $(\tau n, \sigma n)$. Using generating functions
\footnote{We use the following result on bivariate generating functions \cite{Wilf}. Let $a_{i,j}$ be the number of ways of obtaining an outcome $(i,j)\in\mathbb{Z}^2$ in experiment $\cA$, and let $b_{k,l}$ be the number of ways of obtaining an outcome $(k,l)\in\mathbb{Z}^2$ in experiment $\cB$. Also let $c_{p,q}$ be the number of ways of obtaining an outcome $((i,j),(k,l))$ in the combined experiment $(\cA, \cB)$ with sums $i+k=p$ and $j+l=q$. Then the generating functions $A(x,y)=\sum_{i,j} a_{i,j} x^i y^j$, $B(x,y)=\sum_{k,l} b_{k,l} x^k y^l$ and $C(x,y)=\sum_{p,q} c_{p,q} x^p y^q$ are related by $C(x,y) = A(x,y) B(x,y)$.}
, this is given by  
\[
N_{v,t}^{(\gamma n)}(\tau n, \sigma n) = \coeff  \left[ \left( B^{(t)}(x,y) \right) ^{\gamma n}, x^{\tau n} y^{\sigma n} \right]
\]
where $\coeff [p(x,y), x^c y^d ]$ denotes the coefficient of $x^c y^d$ in the bivariate polynomial $p(x,y)$. We make use of the following lemma from \cite[Theorem 2]{Burshtein_Miller}.
\medskip
\begin{lemma}
Let 
\[
B(x,y) = 1 + \sum_{u=1}^{k} \sum_{v=c}^{d} B_{u,v} x^u y^v 
\]
where $k \ge 1$ and $1 \le c \le d$, be a bivariate polynomial satisfying $B_{u,v} \ge 0$ for all $1 \le u \le k$, $c \le v \le d$. For fixed positive rational numbers $\xi$ and $\theta$, consider the set of positive integers $\ell$ such that $\xi \ell \in \mathbb{Z}$, $\theta \ell \in \mathbb{Z}$ and $\coeff [ (B(x,y) ) ^{\ell}, x^{\xi \ell}y^{\theta \ell} ] > 0$. Then either this set is empty, or has infinite cardinality; if $t$ is one such $\ell$, then so is $jt$ for every positive integer $j$. Assuming the latter case, the following limit is well defined and exists:
\begin{multline}
\lim_{\ell\rightarrow \infty} \frac{1}{\ell} \log \coeff \left[ \left( B(x,y) \right) ^{\ell}, x^{\xi \ell}y^{\theta \ell} \right] \\ 
= \max_{\bldsmalleta} \sum_{(i,j) \in S} \eta_{i,j} \log \left( \frac{B_{i,j}}{\eta_{i,j}} \right)
\end{multline}
where $S = \{ (i,j) \in \mathbb{Z}^2 \; : \; B_{i,j} > 0 \}$, $\bldeta = ( \eta_{i,j} )_{(i,j) \in S}$, and the maximization is subject to the constraints 
$\sum_{(i,j) \in S} \eta_{i,j} = 1$, $\sum_{(i,j) \in S} i \eta_{i,j} = \xi$, $\sum_{(i,j) \in S} j \eta_{i,j} = \theta$ and $\eta_{i,j} \ge 0$ for all $(i,j) \in S$.
\label{lemma:optimization_2D}
\end{lemma}
\medskip

Applying this lemma by substituting $B(x,y) = B^{(t)}(x,y)$, $l = \gamma n$, $\xi = \tau/\gamma$ and $\theta = \sigma/\gamma$, we obtain that with $\gamma$ fixed, as $n \rightarrow \infty$ 
\begin{eqnarray}
N_{v,t}^{(\gamma n)}(\tau n, \sigma n) = \coeff \left[ \left( B^{(t)}(x,y) \right) ^{\gamma n}, x^{\tau n} y^{\sigma n} \right] 
\label{eq:Nvt_tau_sigma_start} \\
\rightarrow \exp \left\{ n \gamma \max_{\bldsmalleta^{(t)}} \sum_{(i,j) \in S_t} \eta^{(t)}_{i,j} \log \left( \frac{B^{(t)}_{i,j}}{\eta^{(t)}_{i,j}} \right) \right\} 
\label{eq:Nvt_tau_sigma_mid} \\
\triangleq \exp \left\{ n X^{(\gamma)}_t(\tau, \sigma) \right\}
\label{eq:Nvt_tau_sigma_end}
\end{eqnarray}
where the maximization over $\bldeta^{(t)} = ( \eta^{(t)}_{i,j} )_{(i,j) \in S_t}$ is subject to the constraints $\sum_{(i,j) \in S_t} \eta^{(t)}_{i,j} = 1$, $\sum_{(i,j) \in S_t^{-}} i \eta^{(t)}_{i,j} = \tau / \gamma$, $\sum_{(i,j) \in S_t^{-}} j \eta^{(t)}_{i,j} = \sigma / \gamma$ and $\eta^{(t)}_{i,j} \ge 0$ for all $(i,j) \in S_t$ (recall that the sets $S_t$ and $S_t^{-}$ are given by (\ref{eq:St}) and (\ref{eq:St-})).

\subsection{Growth rate of the weight distribution of the irregular D-GLDPC code ensemble sequence}

Recall that the number of check-valid assignments of weight $\delta m$ is $N_c(\delta m)$; also, the total number of assignments of weight $\delta m$ is $\binom{E}{\delta m}$. Therefore, the probability that a randomly chosen assignment of weight $\delta m$ is check-valid is given by 
\[
P_{\mbox{\scriptsize valid}}(\delta m) = N_c(\delta m) \Big/ \binom{E}{\delta m} \; .
\]
Here we adopt the notation $\delta m = \beta n$; also we have $E = m / \int \rho = n / \int \lambda$. The binomial coefficient may be asymptotically approximated using the fact, based on Stirling's approximation, that as $n \rightarrow \infty$ \cite{Di_Richardson_Urbanke}
\[
\binom{\tau n}{\sigma n} \rightarrow \exp \left\{ n \left[ \sigma \log \left( \frac{e \tau}{\sigma} \right) + 
O(\sigma^2) \right] \right\}
\]
(valid for $0 < \sigma < \tau < 1$) which yields, in this case,
\[
\binom{n / \int \lambda}{\beta n} \rightarrow \exp \left\{ n \left[ \beta \log \left( \frac{e}{\beta \int \lambda} \right) + 
O(\beta^2) \right] \right\}
\]
as $n \rightarrow \infty$. Applying this together with the asymptotic expression (\ref{eq:number_of_check_valid_assign}), and assuming sufficiently small $\beta$, we find that as $n \rightarrow \infty$ (exploiting the fact that $\delta \int \rho = \beta \int \lambda$) 
\begin{equation}
P_{\mbox{\scriptsize valid}}(\beta n) \rightarrow \exp \{ n Y(\beta)\}
\label{eq:Pvalid_limit}
\end{equation}
where 
\[
Y(\beta) = \frac{\beta}{r} \log \left( \frac{e C}{\beta \int \lambda} \right) - \beta \log \left( \frac{e}{\beta \int \lambda} \right) + O(\beta^2) \; .
\]

Next, we note that the expected number of D-GLDPC codewords of weight $\alpha n$ in the ensemble $\cM_n$ is equal to the sum over $\beta$ of the expected numbers of split assignments of split weight $(\alpha n, \beta n)$ which are both check-valid and variable-valid, denoted $N^{v,c}_{\alpha n, \beta n}$:
\[
\mathbb{E}_{\cM_n} \left[ N_{\alpha n} \right] = \sum_{\beta} \mathbb{E}_{\cM_n} [ N^{v,c}_{\alpha n, \beta n} ] \; .
\]
This may then be expressed as  
\begin{multline*}
\mathbb{E}_{\cM_n} \left[ N_{\alpha n} \right] = \\
\sum_{\beta} P_{\mbox{\scriptsize valid}}(\beta n) \sum_{\substack{\sum \alpha_t = \alpha \\ \sum \beta_t = \beta}} \left[ \prod_{t \in I_v} N_{v,t}^{(\delta_t n)}(\alpha_t n, \beta_t n) \right]
\end{multline*}
where the fraction $\delta_t$ of VNs of type $t \in I_v$ is given by (\ref{eq:delta_t_definition})
and the second sum is over all partitions of $\alpha$ and $\beta$ into $n_v$ elements, i.e., we have $\alpha_t, \beta_t \ge 0$ for all $t \in I_v$, and $\sum_{t \in I_v} \alpha_t = \alpha$, $\sum_{t \in I_v} \beta_t = \beta$. 

Now, using (\ref{eq:Nvt_tau_sigma_start})-(\ref{eq:Nvt_tau_sigma_end}), as $n \rightarrow \infty$ we have for each $t \in I_v$
\begin{equation*}
N_{v,t}^{(\delta_t n)}(\alpha_t n, \beta_t n) \rightarrow \exp \left\{ n X^{(\delta_t)}_t(\alpha_t, \beta_t) \right\} \; ,
\end{equation*}
where, for each $t \in I_v$,
\begin{equation}
X^{(\delta_t)}_t(\alpha_t, \beta_t) = \delta_t \max_{\bldsmalleta^{(t)}} \sum_{(i,j) \in S_t} \eta^{(t)}_{i,j} \log \left( \frac{B^{(t)}_{i,j}}{\eta^{(t)}_{i,j}} \right)
\label{eq:X_function}
\end{equation}
and the maximization over $\bldeta^{(t)} = ( \eta^{(t)}_{i,j} )_{(i,j) \in S_t}$ is subject to the constraints 
\begin{equation}
\sum_{(i,j) \in S_t} \eta^{(t)}_{i,j} = 1
\label{eq:eta_sum_constraint}
\end{equation}
\begin{equation}
\sum_{(i,j) \in S_t^{-}} i \eta^{(t)}_{i,j} = \alpha_t / \delta_t
\label{eq:sum_xi_constraint}
\end{equation}
\begin{equation}
\sum_{(i,j) \in S_t^{-}} j \eta^{(t)}_{i,j} = \beta_t / \delta_t
\label{eq:sum_theta_constraint}
\end{equation}
and
\begin{equation}
\eta^{(t)}_{i,j} \ge 0 \quad \forall (i,j) \in S_t \; .
\label{eq:nonnegative_eta_constraint}
\end{equation}
Therefore, recalling (\ref{eq:Pvalid_limit}), we have that as $n \rightarrow \infty$, 
\begin{multline}
\mathbb{E}_{\cM_n} \left[ N_{\alpha n} \right] \rightarrow \\
\sum_{\beta} \sum_{\substack{\sum \alpha_t = \alpha \\ \sum \beta_t = \beta}}
\exp \left\{ n \left[ \sum_{t \in I_v} X^{(\delta_t)}_t(\alpha_t, \beta_t) + Y(\beta) \right] \right\} \; .
\label{eq:sum_of_exp}
\end{multline}
Next, for each $t \in I_v$ we define
\[
F_t(\bldeta^{(t)}) = \eta^{(t)}_{0,0} \log \left( \frac{1}{\eta^{(t)}_{0,0}} \right) - \sum_{(i,j) \in S_t^{-}} \eta^{(t)}_{i,j} \; .
\]
Note that the expression (\ref{eq:sum_of_exp}) is dominated as $n \rightarrow \infty$ by the term which maximizes the argument of the exponential. Thus we may write
\begin{multline}
\!\!\!\!\!\! G(\alpha) = \max_{\beta} \max_{\substack{\sum \alpha_t = \alpha \\ \sum \beta_t = \beta}} \Bigg\{ \sum_{t \in I_v} \delta_t \max_{\bldsmalleta^{(t)}} \Bigg[ \sum_{(i,j) \in S_t^{-}} \eta^{(t)}_{i,j} \log \left( \frac{e B^{(t)}_{i,j}}{\eta^{(t)}_{i,j}} \right) \\
+ F_t(\bldeta^{(t)}) \Bigg] + \frac{\beta}{r} \log \left( \frac{e C}{\beta \int \lambda} \right) \\ - \beta \log \left( \frac{e}{\beta \int \lambda} \right) + O(\beta^2) \Bigg\}
\label{eq:growth_rate_with_O_notation2}
\end{multline}
where the maximization over $\bldeta^{(t)} = ( \eta^{(t)}_{i,j} )_{(i,j) \in S_t^{-}}$ (for each $t \in I_v$) is subject to constraints (\ref{eq:sum_xi_constraint}) and (\ref{eq:sum_theta_constraint}) together with $\eta^{(t)}_{i,j} \ge 0$ for all $(i,j) \in S_t^{-}$.

We next have the following lemma.
\medskip
\begin{lemma}
The expression $\sum_{t \in I_v} \delta_t F_t(\bldeta^{(t)})$ is $O(\alpha^2)$ for any $\bldeta^{(t)}$ satisfying the optimization constraints~(\ref{eq:eta_sum_constraint})-(\ref{eq:nonnegative_eta_constraint})\footnote{Here we use the following standard notation: the real-valued function $f(x)$ is said to be $O (g(x))$ if and only if there exist positive real numbers $k$ and $\epsilon$, both independent of $x$, such that 
\[
\left| f(x) \right| \le k g(x) \quad \forall \; 0 \le x \le \epsilon \; .
\]
}.
\label{lemma:O2_term}
\end{lemma}
\medskip
A proof of this lemma is given in the Appendix. It follows from Lemma \ref{lemma:O2_term} that the expression $\sum_{t \in I_v} \delta_t F_t(\bldeta^{(t)})$ is $O(\alpha^2)$ for the maximizing $\bldeta^{(t)}$. Also, since $\beta/\alpha$ is bounded between two positive constants, any expression which is $O(\beta^2)$ must necessarily also be $O(\alpha^2)$. Therefore
\begin{multline*}
\!\! G(\alpha) = \max_{\beta} \max_{\substack{\sum \alpha_t = \alpha \\ \sum \beta_t = \beta}} \Bigg[ \sum_{t \in I_v} \delta_t \max_{\bldsmalleta^{(t)}} \sum_{(i,j) \in S_t^{-}} \eta^{(t)}_{i,j} \log \left( \frac{e B^{(t)}_{i,j}}{\eta^{(t)}_{i,j}} \right) \\
+ \frac{\beta}{r} \log \left( \frac{e C}{\beta \int \lambda} \right) - \beta \log \left( \frac{e}{\beta \int \lambda} \right) \Bigg] + O(\alpha^2)
\end{multline*}
where the optimization is (as before) subject to the constraints (\ref{eq:sum_xi_constraint}) and (\ref{eq:sum_theta_constraint}) together with $\eta^{(t)}_{i,j} \ge 0$ for all $(i,j) \in S_t^{-}$. In what follows, for convenience of presentation we shall temporarily omit the $O(\alpha^2)$ term in the expression for growth rate.

Next we make the substitution $\gamma^{(t)}_{i,j} = \delta_t \eta^{(t)}_{i,j}$ for all $t \in I_v$, $(i,j) \in S_t^{-}$. This yields
\begin{multline*}
\!\! G(\alpha) = \max_{\beta} \max_{\substack{\sum \alpha_t = \alpha \\ \sum \beta_t = \beta}} \Bigg[ \sum_{t \in I_v} \max_{\bldsmallgamma^{(t)}} \sum_{(i,j) \in S_t^{-}} \gamma^{(t)}_{i,j} \log \left( \frac{e B^{(t)}_{i,j} \delta_t}{\gamma^{(t)}_{i,j}} \right) \\
+ \frac{\beta}{r} \log \left( \frac{e C}{\beta \int \lambda} \right) - \beta \log \left( \frac{e}{\beta \int \lambda} \right) \Bigg]
\end{multline*}
where the maximization over $\bldgamma^{(t)} = ( \gamma^{(t)}_{i,j} )_{(i,j) \in S_t^{-}}$ (for each $t \in I_v$) is subject to the constraints $\sum_{(i,j) \in S_t^{-}} i \gamma^{(t)}_{i,j} = \alpha_t$, $\sum_{(i,j) \in S_t^{-}} j \gamma^{(t)}_{i,j} = \beta_t$, and $\gamma^{(t)}_{i,j} \ge 0$ for all $(i,j) \in S_t^{-}$.
We observe that this maximization may be recast as 
\begin{multline*}
G(\alpha) = \max_{\bldsmallgamma} \Bigg[ \sum_{t \in I_v} \sum_{(i,j) \in S_t^{-}} \gamma^{(t)}_{i,j} \log \left( \frac{e B^{(t)}_{i,j} \delta_t}{\gamma^{(t)}_{i,j}} \right) \\
+ \frac{\beta(\bldgamma)}{r} \log \left( \frac{e C}{\beta(\bldgamma) \int \lambda} \right) - \beta(\bldgamma) \log \left( \frac{e}{\beta(\bldgamma) \int \lambda} \right) \Bigg]
\end{multline*}
where the maximization, which is now over $\bldgamma = (\gamma^{(t)}_{i,j})_{t \in I_v, (i,j) \in S_t^{-}}$, is subject to the constraints 
\[
\sum_{t \in I_v} \sum_{(i,j) \in S_t^{-}} i \gamma^{(t)}_{i,j} = \alpha
\]
and $\gamma^{(t)}_{i,j} \ge 0$ for all $t \in I_v$, $(i,j) \in S_t^{-}$, and where
\[
\beta(\bldgamma) = \sum_{t \in I_v} \sum_{(i,j) \in S_t^{-}} j \gamma^{(t)}_{i,j} \; .
\]
Making the substitution $\nu^{(t)}_{i,j} = \gamma^{(t)}_{i,j} / \alpha$ for all $t \in I_v$, $(i,j) \in S_t^{-}$, we obtain
\begin{multline*}
G(\alpha) = \alpha \max_{\bldsmallnu} \Bigg[ \sum_{t \in I_v} \sum_{(i,j) \in S_t^{-}} \nu^{(t)}_{i,j} \log \left( \frac{e B^{(t)}_{i,j} \delta_t}{\alpha \nu^{(t)}_{i,j}} \right) \\
+ \frac{z(\bldnu)}{r} \log \left( \frac{e C}{\alpha z(\bldnu) \int \lambda} \right) - z(\bldnu) \log \left( \frac{e}{\alpha z(\bldnu) \int \lambda} \right) \Bigg]
\end{multline*}
where the maximization over $\bldnu = (\nu^{(t)}_{i,j})_{t \in I_v, (i,j) \in S_t^{-}}$ is subject to the constraints $\sum_{t \in I_v} \sum_{(i,j) \in S_t^{-}} i \nu^{(t)}_{i,j} = 1$ and $\nu^{(t)}_{i,j} \ge 0$ for all $t \in I_v$, $(i,j) \in S_t^{-}$, and where 
\[
z(\bldnu) = \sum_{t \in I_v} \sum_{(i,j) \in S_t^{-}} j \nu^{(t)}_{i,j} \; .
\]
Assuming the condition $r=2$, we may write 
\[
G(\alpha) = \alpha \max_{\bldsmallnu} \left( K_1(\bldnu) + K_2(\bldnu) \log \alpha \right)
\]
where 
\[
K_2(\bldnu) = \frac{z(\bldnu)}{2} - \sum_{t \in I_v} \sum_{(i,j) \in S_t^{-}} \nu^{(t)}_{i,j} \; .
\]
Next, assuming the condition $p=2$, we make the observation that
\[
z(\bldnu) = \sum_{t \in I_v} \sum_{(i,j) \in S_t^{-}} j \nu^{(t)}_{i,j} \ge 2 \sum_{t \in I_v} \sum_{(i,j) \in S_t^{-}} \nu^{(t)}_{i,j}
\]
with equality if and only if $\nu^{(t)}_{i,j} = 0 $ for all $t \in I_v, (i,j) \in S_t^{-}$ with $j>2$. Therefore $K_2(\bldnu) \ge 0$ with equality if and only if $\nu^{(t)}_{i,j} = 0 $ for all $t \in I_v, (i,j) \in S_t^{-}$ with $j>2$.
Let $\bldnu_1$ and $\bldnu_2$ be two distributions satisfying the optimization constraints, and suppose that $K_2(\bldnu_1) > 0$ and $K_2(\bldnu_2) = 0$. Then for sufficiently small $\alpha$, we must have 
\[
K_1(\bldnu_1) + K_2(\bldnu_1) \log \alpha < K_1(\bldnu_2) + K_2(\bldnu_2) \log \alpha \; .
\]
This follows from the fact that the inequality
\[
K_2(\bldnu_1) \log \alpha < K_1(\bldnu_2) - K_1(\bldnu_1)
\]
will always be satisfied for $\alpha$ sufficiently small (since $\log \alpha \rightarrow~-\infty$ as $\alpha \rightarrow 0$, and recalling that $K_2(\bldnu_1) > 0$).

Therefore, for sufficiently small $\alpha$, the vector $\bldnu$ which maximizes $K_1(\bldnu) + K_2(\bldnu) \log \alpha$ must satisfy $\nu^{(t)}_{i,j} = 0 $ for all $t \in I_v, (i,j) \in S_t^{-}$ with $j>2$. Note that this implies that the maximum, and hence the growth rate, depends only on the check and VNs with minimum distance equal to $2$. Also, recall that for each $t \in X_v$, the set $L_t = \{ i \in \mathbb{Z} \; : \; B^{(t)}_{i,2} > 0 \}$; we contract the vector $\bldnu$ to include only variables $\nu^{(t)}_{i,2}$ where $t \in X_v$ and $i \in L_t$ (since only these may assume positive values).

The growth rate may be written as
\begin{multline}
G(\alpha) = \alpha \max_{\bldsmallnu} \Big[ \sum_{t \in X_v} \sum_{i \in L_t} \nu^{(t)}_{i,2} \log \left( \frac{B^{(t)}_{i,2} \delta_t}{\nu^{(t)}_{i,2}} \right) \\
+ s(\bldnu) \log \frac{s(\bldnu)}{\phi} \Big] \triangleq \alpha \max_{\bldsmallnu} \left( \log R(\bldnu) \right)
\label{eq:log_R_definition}
\end{multline}
where 
\begin{equation}
\phi \triangleq \frac{1}{2 C \int \lambda}
\label{eq:phi_definition}
\end{equation}
and where the function $s(\bldnu)$ is given by
\begin{equation}
s(\bldnu) = \sum_{t \in X_v} \sum_{i \in L_t} \nu^{(t)}_{i,2} \; .
\label{eq:s_nu_definition}
\end{equation}
The maximization over $\bldnu = (\nu^{(t)}_{i,2})_{t \in I_v, i \in L_t}$ in (\ref{eq:log_R_definition}) is subject to the constraints 
\begin{equation}
h(\bldnu) = \sum_{t \in I_v} \sum_{i \in L_t} i \nu^{(t)}_{i,2} = 1
\label{eq:final_sum_constraint}
\end{equation}
and 
\begin{equation}
\nu^{(t)}_{i,2} \ge 0 \quad \forall t \in X_v, i \in L_t \; .
\label{eq:final_inequality_constraint}
\end{equation}
Let the vector $\bldnu$ which maximizes (\ref{eq:log_R_definition}) be denoted by $\tilde{\bldnu}$. Then, our task is to show that 
\[
R(\tilde{\bldnu}) \triangleq \tilde{R} = \frac{1}{P^{-1}(1/C)}
\]
i.e., that 
\begin{equation}
P \left( \frac{1}{\tilde{R}} \right) = \frac{1}{C} \; ,
\label{eq:polynomial_match}
\end{equation}
where the parameter $C$ and the polynomial $P(x)$ are defined in (\ref{eq:C_definition}) and (\ref{eq:Px_definition}), respectively. We show this using Lagrange multipliers, ignoring for the moment the constraint (\ref{eq:final_inequality_constraint}). We have
\[
\frac{\partial \log R(\bldnu)}{\partial \nu^{(t)}_{i,2}} = \log \left( \frac{B^{(t)}_{i,2} \delta_t}{\nu^{(t)}_{i,2}} \right) + \log \left( \frac{s(\bldnu)}{\phi} \right) \:\: ; \:\: \frac{\partial h(\bldnu)}{\partial \nu^{(t)}_{i,2}} = i
\]
so that, at the maximum,
\[
\log \left( \frac{B_{i,2} \delta_t}{\tilde{\nu}^{(t)}_{i,2}} \right) + \log \left( \frac{s(\tilde{\bldnu})}{\phi} \right) = \lambda i
\]
for all $t \in X_v$, $i \in L_t$, and for some Lagrange multiplier $\lambda \in \mathbb{R}$. Substituting back into (\ref{eq:log_R_definition}) and using (\ref{eq:final_sum_constraint}) yields 
\begin{multline}
\log \tilde{R} = \sum_{t \in X_v} \sum_{i \in L_t} \nu^{(t)}_{i,2} \left( \lambda i - \log \left(\frac{s(\tilde{\bldnu})}{\phi} \right) \right) \\ + s(\tilde{\bldnu}) \log \left(\frac{s(\tilde{\bldnu})}{\phi}\right) = \lambda
\end{multline}
i.e., the maximum value of the function $\log R(\bldnu)$ is equal to the Lagrange multiplier. Thus we have
\[
\frac{B^{(t)}_{i,2}}{(\tilde{R})^i} = \left( \frac{\phi}{s(\tilde{\bldnu}) \delta_t} \right) \tilde{\nu}^{(t)}_{i,2}
\]
for all $t \in X_v$, $i \in L_t$. Substituting this into the LHS of (\ref{eq:polynomial_match}) and recalling the definition (\ref{eq:Px_definition}), we obtain
\begin{eqnarray*}
P \left( \frac{1}{\tilde{R}} \right) & = & \sum_{t \in X_v} \sum_{i \in L_t} \frac{2 \lambda_t}{q_t} \frac{B^{(t)}_{i,2}}{(\tilde{R})^i} \\
& = & \sum_{t \in X_v} \sum_{i \in L_t} \frac{\tilde{\nu}^{(t)}_{i,j}}{C s(\tilde{\bldnu})}
= \frac{1}{C}
\end{eqnarray*}
where we have used (\ref{eq:delta_t_definition}), (\ref{eq:phi_definition}) and (\ref{eq:s_nu_definition}). This completes the proof of the theorem. Note that (\ref{eq:growth_rate_case_1}) is a first-order Taylor series expansion around $\alpha = 0$ which directly generalizes the results of \cite{Di_Richardson_Urbanke} and \cite{paolini08:weight} (for irregular LDPC and GLDPC codes respectively) to the case of irregular D-GLDPC codes. Our result indicates that for this case also, the parameter $1 / P^{-1}(1/C)$ plays an analagous role to the parameter $\lambda'(0) \rho'(1)$ for irregular LDPC codes, and to the parameter $\lambda'(0)C$ for irregular GLDPC codes. 

\section{Conclusion}
An expression for the asymptotic growth rate of the weight distribution of D-GLDPC codes for small linear-weight codewords has been derived. The expression assumes the existence of minimum distance $2$ check and variable nodes, and involves the evaluation of a polynomial inverse, derived from the minimum distance $2$ variable nodes, at a point derived from the minimum distance $2$ check nodes. This generalizes known results for LDPC codes and GLDPC codes, and also generalizes the corresponding connection with the stability condition over the BEC.

\appendix

\section*{Proof of Lemma~\ref{lemma:optimization_dominant_term_1D}}
First consider the set of positive rational numbers $\ell$ such that $\xi \ell \in \mathbb{Z}$ and $\coeff ( \{ A(x) \} ^{\ell}, x^{\xi \ell}) > 0$. Then it is easy to see that either this set is empty, or it has infinite cardinality; if $t$ is one such $\ell$, then so is $jt$ for every positive integer $j$ (proof routine by induction). The former case is not of interest to us here. In the latter case, the following limit is well defined and exists \cite[Theorem 1]{Burshtein_Miller}:
\begin{multline}
\lim_{\ell\rightarrow \infty} \frac{1}{\ell} \log \coeff \left[ \left( A(x) \right) ^{\ell}, x^{\xi \ell} \right] \\ 
= \max_{\bldsmallbeta} \sum_{i \in S} \beta_i \log \left( \frac{A_i}{\beta_i} \right)
\end{multline}
where $S = \{ i \in \mathbb{Z} \; : \; A_i > 0 \}$, $\bldbeta = ( \beta_i )_{i \in S}$, and the maximization is subject to the constraints 
\begin{equation}
g(\bldbeta) = \sum_{i \in S} \beta_i = 1
\label{eq:beta_sum_constraint}
\end{equation}
\begin{equation}
h(\bldbeta) = \sum_{i \in S} i \beta_i = \xi
\label{eq:sum_xi_constraint_1D}
\end{equation}
and
\begin{equation}
\beta_i \ge 0 \quad \forall i \in S \; .
\label{eq:nonnegative_beta_constraint}
\end{equation}

We solve this optimization problem using Lagrange multipliers, ignoring for the moment the final constraint. Defining 
\begin{equation}
f(\bldbeta) = \sum_{i \in S} \beta_i \log \left( \frac{A_i}{\beta_i} \right)
\label{eq:f_beta}
\end{equation}
we have 
\[
\frac{\partial f}{\partial\beta_i} = \log \left(\frac{A_i}{e \beta_i}\right) \:\: ; \:\: \frac{\partial g}{\partial\beta_i} = 1 \:\: ; \:\:\: \frac{\partial h}{\partial\beta_i} = i
\]
for all $i \in S$. Therefore we obtain
\begin{equation}
\log \left(\frac{A_i}{e \beta_i}\right) + \lambda + \mu i = 0
\label{eq:Lagrange_1D}
\end{equation} 
for all $i \in S$, where $\lambda$ and $\mu$ are Lagrange multipliers. These equations, together with (\ref{eq:beta_sum_constraint}) and (\ref{eq:sum_xi_constraint_1D}), yield $(|S|+2)$ equations in the $(|S|+2)$ unknowns $\{ \lambda, \mu, \bldbeta \}$. Setting $i = 0$ in (\ref{eq:Lagrange_1D}) yields  
\[
\beta_0 = e^{\lambda-1}
\]
and substituting this back into (\ref{eq:Lagrange_1D}) gives 
\[ 
\beta_i = \beta_0 A_i z^i
\] 
for all $i \in S$, where $z = e^{\mu}$. So from (\ref{eq:sum_xi_constraint_1D})
\[
\beta_0 \sum_{i \in S} i A_i z^i = \xi
\] 
Now for sufficiently small $\xi$, we may approximate
\[
\beta_0 \sum_{i \in S} i A_i z^i \approx \beta_0 c A_c z^c \; .
\]  
Applying this approximation, $\beta_i$ is nonzero only for $i \in \{ 0, c \}$. Therefore, from (\ref{eq:beta_sum_constraint}) and (\ref{eq:sum_xi_constraint_1D}) we obtain the solution
\begin{equation}
\beta_i = \left\{ \begin{array}{cl}
1 - \xi/c & \textrm{if } i = 0 \\
\xi/c & \textrm{if } i = c \\
0 & \textrm{otherwise. }\end{array}\right.
\label{eq:beta_solution}
\end{equation} 
It is easy to see that this solution satisfies (\ref{eq:nonnegative_beta_constraint}). 
Finally, substituting the solution (\ref{eq:beta_solution}) into (\ref{eq:f_beta}) gives 
\begin{eqnarray*}
\max_{\bldsmallbeta} f(\bldbeta) & = & \left(\frac{\xi}{c} - 1 \right) \log \left( 1 - \frac{\xi}{c} \right) + \frac{\xi}{c} \log \left( \frac{c A_c}{\xi} \right) \\
& = & \left(\frac{\xi}{c} - 1 \right) \left( - \frac{\xi}{c} + O(\xi^2) \right) + \frac{\xi}{c} \log \left( \frac{c A_c}{\xi} \right) \\
& = & \frac{\xi}{c} \log \left( \frac{e c A_c}{\xi} \right) + O(\xi^2) \; .
\end{eqnarray*}
This completes the proof of the lemma.

\section*{Proof of Lemma~\ref{lemma:O2_term}}
Consider any $\bldeta^{(t)}$ which satisfies the optimization constraints~(\ref{eq:eta_sum_constraint})-(\ref{eq:nonnegative_eta_constraint}). Since $\alpha = \sum_{t \in I_v} \alpha_t$, $\alpha$ small implies that $\alpha_t$ is small for every $t \in I_v$. From constraint (\ref{eq:sum_xi_constraint}) we conclude that $\eta_{i,j}^{(t)}$ is small for every $t \in I_v$, $(i,j) \in S_t^{-}$, and so $\eta_{0,0}^{(t)}$ is close to $1$ for all $t \in I_v$. Formally, for any $t \in I_v$ the term in the sum over $\bldeta^{(t)}$ in (\ref{eq:X_function}) corresponding to $(i,j) = (0,0)$ may be written as (here we use (\ref{eq:eta_sum_constraint}), and the Taylor series of
$\log\left(1-x\right)$ around $x=0$)
\begin{multline*}
\eta^{(t)}_{0,0} \log \left( \frac{1}{\eta^{(t)}_{0,0}} \right) = \Big( \sum_{(i,j) \in S_t^{-}} \eta^{(t)}_{i,j} - 1 \Big) \log \Big( 1 - \sum_{(i,j) \in S_t^{-}} \eta^{(t)}_{i,j} \Big) \\
= \Big( \sum_{(i,j) \in S_t^{-}} \eta^{(t)}_{i,j} - 1 \Big) \Bigg(- \sum_{(i,j) \in S_t^{-}} \eta^{(t)}_{i,j} + O \Big( \Big( \sum_{(i,j) \in S_t^{-}} \eta^{(t)}_{i,j} \Big)^2 \Big) \Bigg) \\
= \sum_{(i,j) \in S_t^{-}} \eta^{(t)}_{i,j} + O \Big( \Big( \sum_{(i,j) \in S_t^{-}} \eta^{(t)}_{i,j} \Big)^2 \Big) 
\end{multline*}
Therefore we have
\begin{equation} 
\left| F_t(\bldeta^{(t)}) \right| \le k_t \Big( \sum_{(i,j) \in S_t^{-}} \eta^{(t)}_{i,j} \Big)^2
\label{eq:F_t_bound}
\end{equation}
for some $k_t > 0$ independent of $\{ \eta^{(t)}_{i,j} \}_{(i,j) \in S_t^{-}}$. It follows that
\begin{equation}
\left| \sum_{t \in I_v} \delta_t F_t(\bldeta^{(t)}) \right| \le \sum_{t \in I_v} \delta_t \left| F_t(\bldeta^{(t)}) \right| \le \sum_{t \in I_v} \delta'_t \Big( \sum_{(i,j) \in S_t^{-}} \eta^{(t)}_{i,j} \Big)^2
\label{eq:F_t_ineq_1}
\end{equation} 
where $\delta'_t = k_t \delta_t$ for each $t \in I_v$. Also, by (\ref{eq:sum_xi_constraint}) we have $\sum_{(i,j) \in S_t^{-}} \eta^{(t)}_{i,j} \le \alpha_t / \delta_t$ and therefore
\begin{equation}
\sum_{t \in I_v} \delta'_t \Big( \sum_{(i,j) \in S_t^{-}} \eta^{(t)}_{i,j} \Big)^2 \le \sum_{t \in I_v} \left( \frac{\delta'_t}{\delta_t^2} \right) \alpha_t^2
\label{eq:F_t_ineq_2}
\end{equation} 
Denote $\delta = \max_{t \in I_v} \{ \delta'_t / \delta_t^2 \}$; then, combining (\ref{eq:F_t_ineq_1}) and (\ref{eq:F_t_ineq_2}), 
\[
\left| \sum_{t \in I_v} \delta_t F_t(\bldeta^{(t)}) \right| \le \delta \sum_{t \in I_v} \alpha_t^2 < \delta \left( \sum_{t \in I_v} \alpha_t \right) ^2 = \delta \alpha^2
\]
and thus the expression $\sum_{t \in I_v} \delta_t F_t(\bldeta^{(t)})$ is $O(\alpha^2)$, as desired.

\section*{Acknowledgments}
This work was supported in part by the EC under Seventh FP grant agreement ICT OPTIMIX n. INFSO-ICT-214625 and in part by the University of Bologna (ISA-ESRF fellowship). 


\bibliographystyle{IEEEtran}

\end{document}